\documentclass[english,pra, twocolumn, showpacs, showkeyw]{revtex4-1}
\usepackage[T1]{fontenc}
\usepackage[latin9]{inputenc}
\usepackage{amsmath}
\usepackage{graphicx}

\makeatletter
\usepackage[colorlinks = true,
            linkcolor =red,
            urlcolor  =magenta,
            citecolor = blue,
            anchorcolor = blue]{hyperref}

\makeatother

\usepackage{babel}
\begin{document}
\title{Theoretical investigation of the relations between quantum decoherence
and weak-to-strong measurement transition}
\author{Xiao-Feng Song, Shuang Liu, Xi-Hao Chen}
\author{Yusuf Turek}
\email{yusufu1984@hotmail.com}

\affiliation{$^{1}$School of Physics, Liaoning University, Shenyang, Liaoning
110036, China}
\date{\today}
\begin{abstract}
This paper delves into the crucial aspects of pointer-induced quantum
decoherence and the transition between von Neumann's projective strong
measurement and Aharonov's weak measurement. Both phenomena significantly
impact the dynamical understanding of quantum measurement processes.
Specifically, we focus on the interplay between quantum decoherence
and the transition from weak to strong measurement by deducing and
comparing the quantum decoherence and weak-to-strong measurement transition
factors within a general model and using the well-known Stern-Gerlach
experiment as an illustrative example. Our findings reveal that both
phenomena can be effectively characterized by a universal transition
factor intricately linked to the coupling between the system and the
measurement apparatus. The analysis presented can clarify the mechanism
behind the relations of quantum decoherence to the weak measurement
and weak-to-strong measurement transition.
\end{abstract}
\maketitle

\section{\label{sec:1}Introduction}

Measurements in quantum mechanics have posed a longstanding and formidable
challenge, playing a fundamental role in exploring the properties
of quantum systems \citep{Peres1995-PEREQT,Aharonov2005}. It is widely
recognized that von Neumann developed the first model to describe
strong quantum measurements by treating both the system under test
and the quantum measuring instrument, with strong interactions between
them \citep{Edwards1955}. The well-known Stern-Gerlach (SG) experiment
is a typical model of quantum strong measurements, which can be interpreted
as a quantum measurement process that measures the spin of the particles
through their spatial distribution \citep{Sakurai2020}. During the
strong measurement process, the system's state collapses to one of
its eigenstates due to the significant interaction between the system
and the measuring instrument. This process is advantageous because
it allows the acquisition of the desired system information through
a single measurement. However, the wave packet collapse induced by
strong measurement is irreversible, suggesting that the measured quantum
state is unlikely to revert to its original state.

The issues associated with strong quantum measurement have been extensively
elucidated within the framework of quantum mechanics, focusing on
the interaction between the measuring device and the system \citep{Schlosshauer2008DecoherenceAT}.
One of these important issues is quantum decoherence, whereby the
measurement devices disrupt over time the quantum coherence of superpositions
\citep{2007Breuer}. In 1970, Zeh \citep{Zeh1970-ZEHOTI} authored
the first paper on decoherence, highlighting that realistic macroscopic
quantum systems are inherently open, undergoing strong interactions
with their environments. However, a crucial advancement in decoherence
occurred in the 1980s with the introduction of the term decoherence.
This milestone was notably propelled by the seminal contributions
of Zurek \citep{PhysRevD.24.1516,PhysRevD.26.1862}, who underscored
the paramount significance of preserving quantum correlations, establishing
it as a pivotal criterion for discerning preferred states within the
decoherence framework \citep{Joos:1984uk,Zurek1986}. The reader is
referred to recent reviews in the field for further details of quantum
decoherence \citep{RevModPhys.75.715,RevModPhys.76.1267,2007B,20191}. 

As previously discussed, when the interaction between the measured
system and the measuring apparatus is intense, decoherence takes place.
This results in the collapse of the measured system into its corresponding
eigenstate of the measured observable, imposing information loss.
Moreover, the irreversible nature of strong measurements implies that
the measured quantum state is unlikely to return to its original state.
Thus, the weak measurement term was introduced to address the challenges
of strong quantum measurement. In weak measurement, the coupling between
the system and the measuring apparatus is minimal, thereby avoiding
wave function collapse \citep{Tollaksen2010}. The measurement value
is obtained by incorporating a suitable post-selection step in the
weak measurement process, often termed the \textit{weak value}, which
may fall beyond the observable's eigenvalue spectrum. This leads to
a phenomenon known as \textit{weak value amplification} (WVA), which
has proven beneficial for detecting and examining minute effects within
linear optical systems \citep{Zhou2013WeakvalueAO,PhysRevA.82.011802,PhysRevLett.112.200401,articleViza}.
It has also assisted in exploring quantum mechanics and its applications
\citep{70,RN2035,RN2066,RevModPhys.86.307}. 

Despite the successful explanations of various quantum measurement
phenomena achieved by both the theory of strong and weak measurements,
an unavoidable question naturally arises regarding the feasibility
of transitioning from weak to strong measurement by modifying the
interaction between the system and the apparatus \citep{Ferraioli2019-FERTMP}.
Transitioning from weak to strong measurement can be traced back to
the investigation of Zhu et al. \citep{PhysRevA.84.052111}. Their
study examined quantum measurements involving pre-selection and post-selection
and studied the pointer position and momentum shifts without relying
on approximations, thus broadening the scope to encompass strong interactions.
Ban's research \citep{Ban} focused on exploring whether another form
of observable average existed distinct from weak and strong values
in a post-selected quantum system. The author's findings reveal that
if eigen-projectors of a measured observable solely represent the
measurement's impact on the system, the conditional average is a combination
of strong and weak observable values. Later, several studies explored
the transition from weak to strong measurement by selecting different
pointers. For instance, Pan et al. \citep{Nature2020} designed a
system to experimentally observe the transition from weak to strong
measurement in a Gaussian state by modulating a global transition
factor. Orszag et al. \citep{ArayaSossa2021InfluenceOS} investigated
the measurement transition for a coherent squeezed pointer state.
Their study demonstrates a pathway from weak to strong measurements
while preserving the constancy of the global transition factor. This
methodology offers an alternative pathway for exploring the measurement
transition. Furthermore, a general approach addressing the transition
from weak to strong measurement employed Fock state-based states as
the pointer state. This approach is based on the principle that the
Hermitian nature of the photon number operator allows for any state
to be expanded based on $\left|n\right\rangle $ \citep{TUREK2023128663}.

Although quantum decoherence and weak-to-strong measurement transitions
have been extensively studied, their potential relationship remains
unexplored. Hence, this paper investigates the connection between
these two phenomena to uncover the underlying physical implications
by employing a generic computational model for each phenomenon. Surprisingly,
our analytic results reveal that the quantum decoherence and transition
factors share a common mathematical form. Besides, we validate our
findings further by applying this model to the renowned SG experiment
and obtain consistent outcomes for both factors. Moreover, upon separately
analyzing the two asymptotic states with factor values approaching
0 and 1, we observe that the measured system demonstrates consistent
trends under identical factors, whether within the context of decoherence
or the transition from weak to strong measurements.

The remainder of this paper is organized as follows. Section. \ref{sec:2}
briefly introduces the quantum decoherence and weak-to-strong measurement
transition models. Section. \ref{sec:3} discusses the dynamical measurement
process of the SG experiment as a typical example of our proposal.
Section. \ref{sec:4} discusses our findings and concludes this work.

\section{\label{sec:2} Brief description of quantum decoherence and weak-to-strong
measurement transition}

\subsection{\label{subsec:1} Quantum decoherence and decoherence function }

Any standard quantum measurement model has two components: system
(measured system) and pointer (measurement device/measuring system/measurement
apparatus). Its Hamiltonian is formulated as follows,

\begin{equation}
H=H_{s}+H_{p}+H_{I}=H_{0}+H_{I}\text{.}\label{eq:1}
\end{equation}
 In the equation above, $H_{s}$ and $H_{p}$ represent the Hamiltonian
of the system and measurement pointer, respectively, and $H_{I}$
describes the interaction between the system and the pointer. It is
important to note that this work solely focuses on the ideal measurement
model without considering the noise caused by a reservoir. In general,
the interaction Hamiltonian $H_{I}$ takes the von Neumann measurement
form as

\begin{equation}
H_{I}=gA\text{\ensuremath{\otimes Q}},\ \ \label{eq:2-1}
\end{equation}
 where $A$ is the system observable we want to obtain information
from the measurement model and $Q=Q^{\dagger}$ is an arbitrary pointer
operator. The $Q$ is usually the position ($X$) or momentum ($P$)
operator, easing the system information acquisition in the lab. The
interaction strength $g$ between the system and the pointer is usually
an impulsive function, which is only effective over a very short time
interval to guarantee the precision of the measurement result. We
assume that the system Hamiltonian commutes with the observable $A$,
which yields,
\begin{equation}
[H,A]=[H_{I},A]=[H_{s},A]=0,\label{eq:3}
\end{equation}
such that the system observable $A$ involves conserved quantities.
As a consequence, the mean energy of the system is constant in time,
i.e.,
\begin{equation}
\frac{d}{dt}\langle H_{s}(t)\rangle=0.\label{eq:4}
\end{equation}
From the above assumption, the system and observable $A$ have the
following eigenvalue function, i.e.,
\begin{equation}
H_{s}\vert a_{i}\rangle=E_{i}\vert a_{i}\rangle,\ \ \ A_{i}\vert a_{i}\rangle=a_{i}\vert a_{i}\rangle.
\end{equation}
Initially, the system and the pointer are relatively independent and
the initial state of the composite system is written as 
\begin{equation}
\vert\Psi(0)\rangle=\vert\psi_{i}\rangle\otimes\vert\phi\rangle.\label{eq:6-2}
\end{equation}
 Here $\vert\psi_{i}\rangle=\sum_{i}\alpha_{i}\vert a_{i}\rangle$
with $\alpha_{i}=\langle a_{i}\vert\psi_{i}\rangle$ and $\phi$ representing
the initial states of the system and the pointer, respectively. The
initial system state $\vert\psi_{i}\rangle$ can be expressed in terms
of density matrix as 
\begin{align}
\rho_{s}(0) & =\vert\psi_{i}\rangle\langle\psi_{i}\vert\nonumber \\
 & =\sum_{i}\vert\alpha_{i}\vert^{2}\vert a_{i}\rangle\langle a_{i}\vert+\sum_{j,i}\alpha_{i}\alpha_{j}^{\ast}\vert a_{i}\rangle\langle a_{j}\vert,
\end{align}
 where $\vert\alpha_{i}\vert^{2}$ is the measuring probability of
eigenvalue $a_{i}$ of the observable $A$ corresponding to the eigenstate
$\vert a_{i}\rangle$ if the system is prepared in $\vert\psi_{i}\rangle$.
In order to obtain a determined outcome, the second term (off-diagonal,
the part that represented the coherence) of $\rho$ has to vanish
after the measurement. This means that, after the measurement, the
system is a mixture of the eigenstates of the measured observable.
Next, we describe this dynamic transition process.

\begin{figure}
\includegraphics[width=8cm,height=6cm]{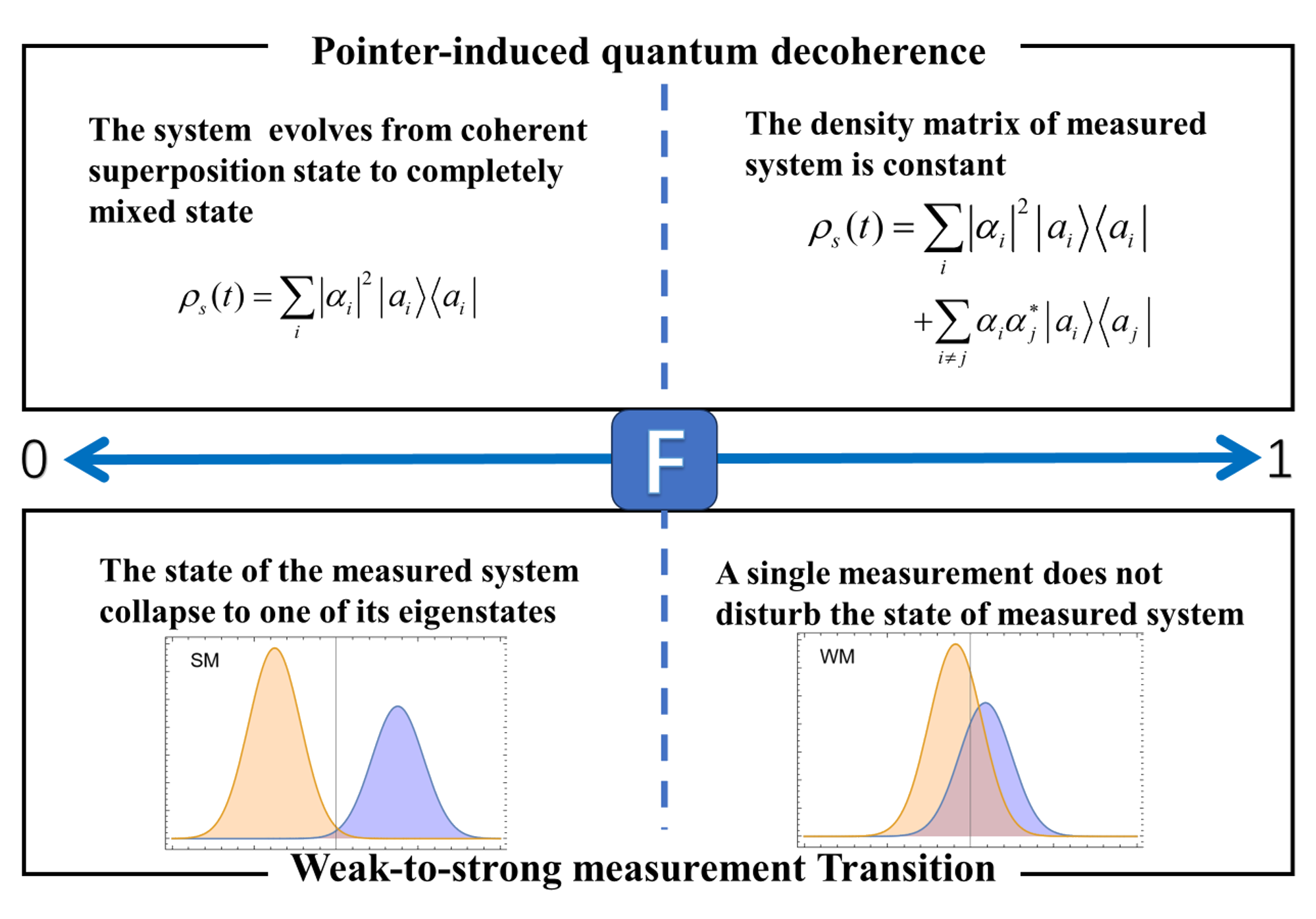}

\caption{\label{fig:1}Schematic relations between decoherence and weak-to-strong
measurement transition model. }

\end{figure}

After completing measurements, the time evolution of the total system
is characterized by
\begin{equation}
\vert\Psi(t)\rangle=U\left(t\right)\vert\Psi(0)\rangle,\label{eq:7-1}
\end{equation}
where $U(t)$ is a unitary time evolution operator defined as
\begin{align}
U(t) & =\exp\left[-iHt\right]=\exp\left[-i\left(H_{s}+H_{p}+H_{I}\right)t\right]\nonumber \\
 & =\sum_{i}e^{-iE_{i}t}e^{-i(H_{p}+g_{0}a_{i}Q)t}\vert a_{i}\rangle\langle a_{i}\vert.\label{eq:8}
\end{align}
By substituting Eq. (\ref{eq:8}) into Eq. (\ref{eq:7-1}) we obtain

\begin{equation}
\vert\Psi(t)\rangle=\sum_{i}\alpha_{i}e^{-iE_{i}t}\vert a_{i}\rangle\vert\phi_{i}(t)\rangle.\label{eq:9-1}
\end{equation}
 Here $\vert\phi_{i}(t)\rangle=e^{-i(H_{p}+g_{0}a_{i}Q)t}\vert\phi\rangle$
represents the final states of the pointer and contains the information
of the system observable $A$. $\vert\Psi(t)\rangle$ indicates that
the system and pointer become entangled after time evolution and cannot
be separated. However, we can determine the system state by tracing
out the degrees of freedom of the pointer as
\begin{align}
\rho_{s}(t) & =Tr_{p}\left(\vert\Psi(t)\rangle\langle\vert\Psi(t)\right)\nonumber \\
 & =\sum_{i}\vert\alpha_{i}\vert^{2}\vert a_{i}\rangle\langle a_{i}\vert\nonumber \\
 & +\sum_{i\neq j}\alpha_{j}^{\ast}\alpha_{i}e^{-i(E_{i}-E_{j})t}\vert a_{i}\rangle\langle a_{j}\vert\langle\phi_{j}(t)\vert\phi_{i}(t)\rangle.\label{eq:10}
\end{align}
 The above expression reveals that the diagonal terms of $\rho_{s}(t)$
remain unchanged with time, while the off-diagonal terms vary over
time. The dependence of matrix element $\left\langle a_{i}\right|\rho_{s}(t)\left|a_{j}\right\rangle $
on time is given in the form of overlapping integrals of $\left|\phi_{i}(t)\right\rangle $
and $\left|\phi_{j}(t)\right\rangle $, and the impact exerted by
the pointer (measuring apparatus) on the statistical measurement outcomes
is effectively subsumed in the overlap. The amount of overlap is a
quantitative measure delineating the degree of interference. Generally,
\begin{equation}
F(t)=\left|\left\langle \phi_{i}(t)\mid\phi_{j}(t)\right\rangle \right|=exp\left[-\Gamma_{ij}(t)\right],\Gamma_{ij}(t)\ge0.\ \label{eq:11-1}
\end{equation}
 The above formula describes the behavior of the non-diagonal elements
of the reduced density matrix $\rho_{s}(t)$ when $i\ne j$. Its time
dependence is related to many elements, such as the specific form
or the system-pointer coupling, on the underlying model's various
parameters and the initial state's properties. Therefore, $F(t)$
is called the \textit{decoherence function}. For many physical systems,
the irreversible dynamics induced by the system-pointer (system-reservoir)
interaction rapidly decreases the overlap $\langle\phi_{j}(t)\vert\phi_{i}(t)\rangle$
when $i\ne j$. Thus, to quantitatively describe the decreasing $F(t)$,
we introduce the \textit{decoherence time} $\tau_{D}$ term. If for
$i\neq j$ the overlap of the states $\vert\phi_{i}(t)\rangle$ and
$\left|\phi_{j}(t)\right\rangle $ approaches to zero after large
times compared to $\tau_{D}$ such that 
\begin{equation}
\langle\phi_{j}(t)\vert\phi_{i}(t)\rangle\rightarrow\delta_{ij},\ \ for\ \ t\gg\tau_{D},
\end{equation}
then, the reduced density matrix of the system becomes as
\begin{equation}
\rho_{s}(t)\to\sum\limits _{i}\left|\alpha_{i}\right|^{2}\left|a_{i}\right\rangle \left\langle a_{i}\right|.\label{eq:12-1}
\end{equation}
 This result shows that the coherence of our system's density matrix
of our system vanishes after a long time ($t\gg\tau_{D}$) interaction
with the pointer. In the measurement problem, after $t\gg\tau_{D}$
the state $\rho_{s}(t)$ of the system behaves as an incoherent mixture
of the state $\vert a_{i}\rangle$, so that the interference terms
of the form $\langle a_{i}\vert A\vert a_{j}\rangle$ ($i\neq j$)
of any system observable $A$ no longer occurs in the expectation
value. In other words, after a longer time of interaction between
the system and the pointer or environment, the super-positions of
the states $\vert a_{i}\rangle$ are destroyed locally, meaning they
are unobservable for all measurements executed exclusively on the
system. The dynamic transition process expressed by Eq. (\ref{eq:12-1})
is called \textit{decoherence}. The main idea of our proposal to reveal
the relations between quantum decoherence and weak-to-strong measurement
transition is illustrated in Fig. (\ref{fig:1}).

\subsection{\label{subsec:2} A model of pointer-induced decoherence}

This subsection introduces a pointer-induced decoherence model. As
mentioned previously, in the ideal measurement, the Hamiltonian part
only adds a phase factor to the total system state after time evolution
and does not affect measurement results. Thus, we assume the Hamiltonian
of one measurement as 
\begin{equation}
H=\frac{p^{2}}{2m}-g(t)x\otimes A.\label{eq:9}
\end{equation}
 Here, $A=\sum_{i}a_{i}\vert a_{i}\rangle\langle a_{i}\vert$ as defined
is a system observable and $m$ is the mass of a quantum object. The
initial state of the total system is $\vert\psi_{i}\rangle\otimes\vert\phi\rangle$,
where $\vert\psi_{i}\rangle$ as defined in Sec. ( \ref{subsec:1})
and the pointer is assumed to be a Gaussian profile as
\begin{align}
\vert\phi\rangle & =\left(\frac{1}{2\pi\sigma^{2}}\right)^{\frac{1}{4}}\int\exp(-\frac{x^{2}}{4\sigma^{2}})dx\vert x\rangle,\label{eq:11}
\end{align}
 where $\sigma$ is the width of the Gaussian wave packet. The evolution
of the total system can be written as 
\begin{equation}
\vert\Psi(t)\rangle=U(t)\vert\psi_{i}\rangle\otimes\vert\phi\rangle\label{eq:12}
\end{equation}
 with $U(t)=\exp\left(-iHt\right)$. The explicit form of $\vert\Psi(t)\rangle$
is obtained by factorizing the unitary operator $U(t)$, accomplished
by adopting the Wei-Norman technique \citep{Sun_1991} as
\begin{equation}
U(t)=\sum_{i}e^{g_{1}}e^{g_{2}p^{2}}e^{g_{3}p}e^{g_{4}x}\vert a_{i}\rangle\langle a_{i}\vert\label{eq:13-1}
\end{equation}
with 
\begin{align}
g_{1} & =-\frac{i(ga_{i})^{2}}{6m}t^{3},\\
g_{2} & =-\frac{it}{2m},\\
g_{3} & =\frac{it^{2}ga_{i}}{2m},\\
g_{4} & =iga_{i}t.
\end{align}
 By substituting Eq. (\ref{eq:13-1}) into Eq. (\ref{eq:12}), we
obtain
\begin{equation}
\vert\Psi(t)\rangle=\sum_{i}\alpha_{i}\vert a_{i}\rangle\vert\phi_{i}(t)\rangle.\label{eq:18}
\end{equation}
 Here, $\vert\phi_{i}(t)\rangle$ are final states of the pointer,
and reads as 
\begin{align}
\vert\phi_{i}(t)\rangle & =\frac{\left(\sigma^{2}/2\pi\right)^{1/4}}{\sqrt{\sigma^{2}+\frac{it}{2m}}}e^{-i\theta(t)}e^{-iga_{i}tx}\nonumber \\
 & \times\exp\left[-\frac{\left(x-\frac{ga_{i}t^{2}}{2m}\right)^{2}}{4(\sigma^{2}+\frac{it}{2m})}\right]\label{eq:19}
\end{align}
with $\theta(t)=\frac{\left(ga_{j}\right)^{2}}{6m}t^{3}$. The expression
above represents a Gaussian wavepacket with width $\sigma(t)$=$\sigma\left(1+\frac{t^{2}}{4m^{2}\sigma^{4}}\right)^{1/2}$and
a central position at $x_{i}=\frac{ga_{i}t^{2}}{2m}$. 

The final state of the system after time evolution is 
\begin{align}
\rho_{s}^{\prime}(t) & =Tr_{p}\left(\vert\Psi(t)\rangle\langle\Psi(t)\vert\right)\nonumber \\
 & =\sum_{i}\vert\alpha_{i}\vert^{2}\vert a_{i}\rangle\langle a_{i}\vert+\sum_{i\neq j}\alpha_{i}\alpha_{j}^{\ast}\vert a_{i}\rangle\langle a_{j}\vert F_{ij}.\label{eq:20}
\end{align}
Here, the decoherence factor is given by
\begin{align}
F & =\vert F_{ij}\vert=\vert\langle\phi_{i}(t)\vert\phi_{j}(t)\rangle\vert\nonumber \\
 & =\exp\left[-\frac{5}{8}\frac{(\triangle x)^{2}}{\sigma^{2}(t)}-\frac{t^{2}}{32\sigma^{4}m^{2}}\frac{(\triangle x)^{2}}{\sigma^{2}(t)}-\frac{2\sigma^{4}m^{2}(\triangle x)^{2}}{t^{2}\sigma^{2}(t)}\right],\label{eq:21}
\end{align}
where $\triangle x=gt^{2}(a_{i}-a_{j})/2m$. This function can quantify
the degree of decoherence as a function of time $t$. It should be
noted that the distinguishability condition of the wavepackets at
time $t$ is the distance of the center of two near wavepackets larger
than its width, i.e., $\triangle x\gg\sigma(t)$ . This condition
can easily be satisfied if time $t$ is long enough.

\subsection{\label{subsec:3}Weak-to-strong measurement transition model}

Let the weak-to-strong measurement transition model have the same
Hamiltonian as in the above subsection. Using the pointer shift, we
read the system's observable values in all measurement schemes. Therefore,
three value types of the observable $A$ correspond to different measurement
circumstances. To clearly understand the mechanism of measurement
transition, let's first briefly introduce each value of $A$ :

1.\textsl{ Expectation value}. Suppose a system state $\vert\psi_{i}\rangle=\sum_{j}\alpha_{j}\vert a_{j}\rangle$
with $\sum_{j}\vert\alpha_{j}\vert^{2}=1$, then the expectation value
of the observable under the state $\vert\psi_{i}\rangle$ is given
by
\begin{equation}
\langle A\rangle=\sum_{j}a_{j}\vert\alpha_{j}\vert^{2}.\label{eq:6-1}
\end{equation}
 This expectation value can obtained by reading the position shift
$\delta x$ under the state given in Eq. (\ref{eq:18}), i.e.,
\begin{equation}
\delta x=\langle\Psi(t)\vert X\vert\Psi(t)\rangle-\langle\phi\vert X\vert\phi\rangle=\frac{gt^{2}}{2m}\langle A\rangle.\label{eq:28-1}
\end{equation}

2. \textsl{Conditional Expectation value}. In time-symmetric quantum
mechanics \citep{PhysRev.134.B1410,Aharonov1991CompleteDO,Aharonov2008},
if we take a post-selection by using the state $\vert\psi_{f}\rangle=\sum_{j}\beta_{j}\vert a_{j}\rangle$
with $\sum_{j}\vert\beta_{j}\vert^{2}=1$ after some evolution, the
conditional expectation value of the observable $A$ is determined
by \citep{PhysRev.134.B1410}
\begin{align}
\langle A\rangle_{c} & =\frac{\sum_{j}a_{j}\vert\langle\psi_{f}\vert a_{j}\rangle\langle a_{j}\vert\psi_{i}\rangle\vert^{2}}{\sum_{j}\vert\langle\psi_{f}\vert a_{j}\rangle\langle a_{j}\vert\psi_{i}\rangle\vert^{2}}\nonumber \\
 & =\frac{\sum_{j}a_{j}\vert\alpha_{j}\beta_{j}^{\ast}\vert^{2}}{\sum_{j}\vert\alpha_{j}\beta_{j}^{\ast}\vert^{2}}.\label{eq:6}
\end{align}
This value is also called the post-selected strong value of $A$.
The above processes we assume that the coupling between the measured
system and the pointer is strong enough so that the spatial sub-wave-packets
of the pointer corresponding to the different eigenvalues of the observable
are distinguishable, i.e., $g\triangle a\gg\sigma$. Here, $\triangle a=a_{i}-a_{i-1}$
and $\sigma$ represent the differences between neighboring eigenvalues
and the width of the sub-wave packets, respectively. 

\begin{figure}
\includegraphics[width=8cm]{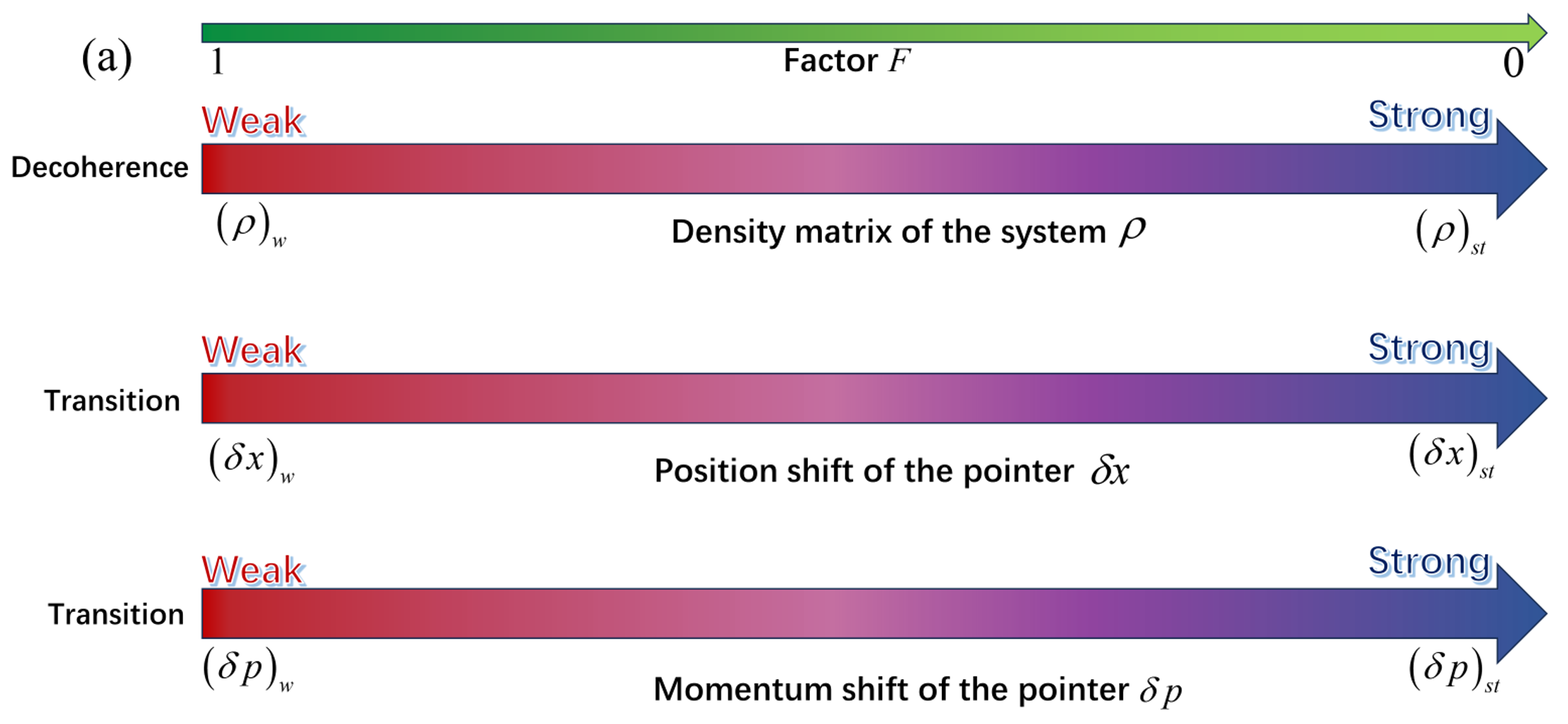}

\vspace{2cc}

\includegraphics[width=8cm]{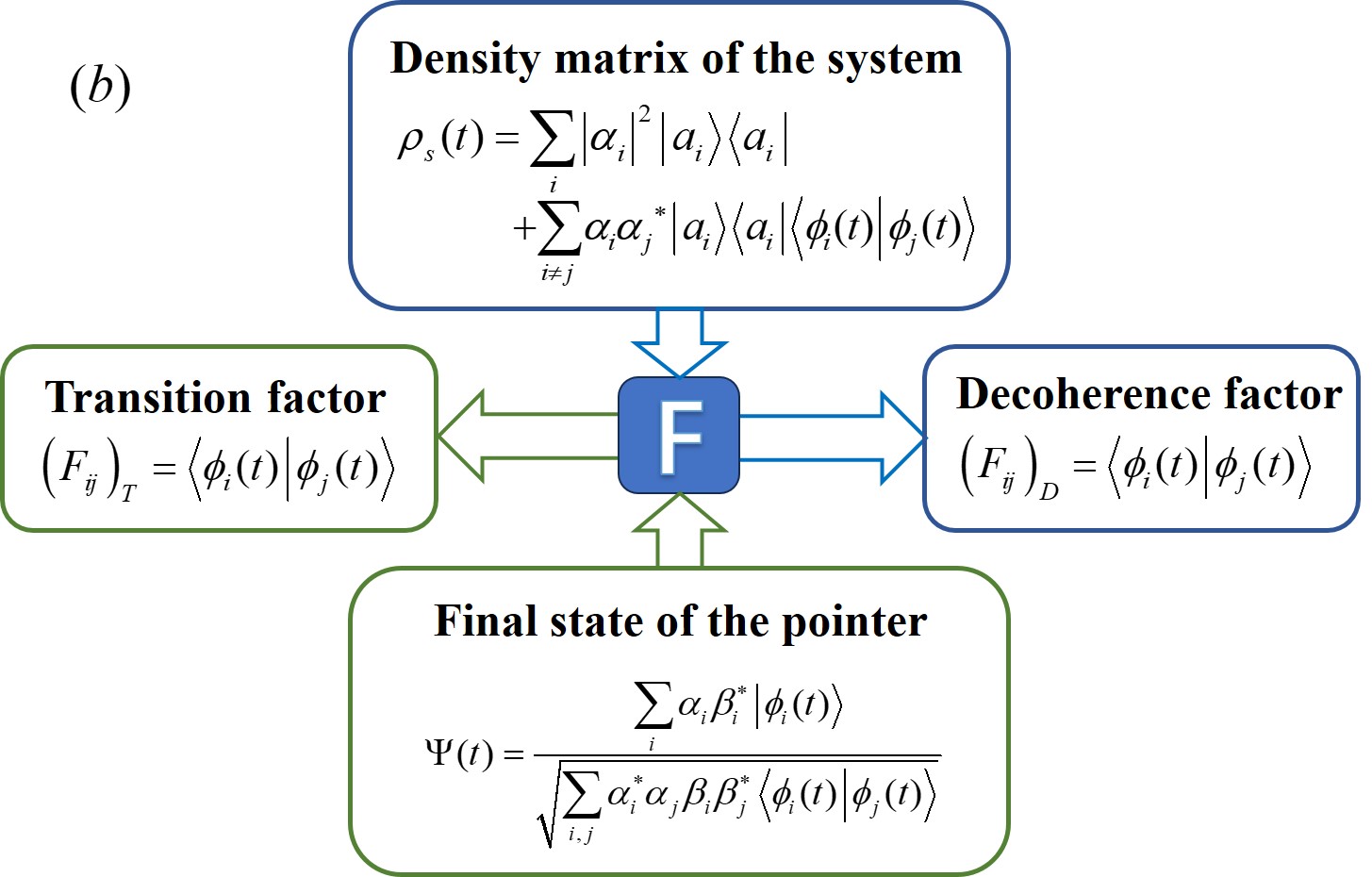}

\caption{\label{fig:2}Weak-to-strong measurement model and its relation with
the quantum decoherence factor. (a) measurement readout by the displacements
of position and momentum observables. (b) decoherence and transition
factors describing different physical processes but share the same
expression.}

\end{figure}

3. \textsl{Weak value}. The weak value of observable $A$ with pre-
and post-selected state reads as 
\begin{align}
\langle A\rangle_{w} & =\frac{\langle\psi_{i}\vert A\vert\psi_{f}\rangle}{\langle\psi_{i}\vert\psi_{f}\rangle}=\frac{\sum_{j}\sum_{k}\alpha_{j}\beta_{k}^{\ast}\langle a_{k}\vert A\vert a_{j}\rangle}{\sum_{j}\sum_{k}\alpha_{j}\beta_{k}^{\ast}\langle a_{k}\vert a_{j}\rangle}\nonumber \\
 & =\frac{\sum_{j}a_{j}\alpha_{j}\beta_{j}^{\ast}}{\sum_{j}\alpha_{j}\beta_{j}^{\ast}}.\label{eq:13}
\end{align}
It can be seen that, in general, the conditional expectation value
$\langle A\rangle_{c}$ and the weak value $\langle A\rangle_{w}$
are different {[}please see the Eq. (\ref{eq:6}) and Eq. (\ref{eq:13})
{]}, and correspond to different measurement strengths. Hence, the
(conditional) expectation value of the system observable is related
to the (post-selected) strong measurement, while the post-selected
weak measurement causes the weak value. If $\beta_{i}=\alpha_{i},$
the above-introduced conditional expectation value and weak value
are reduced to the typical expectation value of the observable $A$
as given in Eq. (\ref{eq:6-1}). Actually, the $\langle A\rangle_{w}$
and $\langle A\rangle_{c}$ are the two extreme values of the transition
value of observable $A$ defined below
\begin{align}
A_{T} & =\frac{\langle\psi_{f}\vert A\rho_{s}^{\prime}(t)\vert\psi_{f}\rangle}{\langle\psi_{f}\vert\rho_{s}^{\prime}(t)\vert\psi_{f}\rangle}\nonumber \\
 & =\frac{\sum_{i}a_{i}\vert\alpha_{i}\beta_{i}\vert^{2}+\sum_{i\neq j}a_{i}\beta_{j}\beta_{i}^{\ast}\alpha_{i}\alpha_{j}^{\ast}\langle\phi_{i}(t)\vert\phi_{j}(t)\rangle}{\sum_{i}\vert\alpha_{i}\beta_{i}\vert^{2}+\sum_{i\neq j}\beta_{j}\beta_{i}^{\ast}\alpha_{i}\alpha_{j}^{\ast}\langle\phi_{i}(t)\vert\phi_{j}(t)\rangle}.\label{eq:31-1}
\end{align}
 The expression above reveals that the transition value depends on
the overlap $\langle\phi_{i}(t)\vert\phi_{j}(t)\rangle$ of the states
$\vert\phi_{i}(t)\rangle$ and $\left|\phi_{j}(t)\right\rangle $,
and its module $F$ is decoherence function of the system {[}see Eq.
(\ref{eq:21}){]}. $F$ is an exponentially decreasing function of
time $t$ and coupling strength $g$. If $gt^{2}\vert a_{i}-a_{j}\vert/2m\gg\sigma(t)$,
then $F$ approaches zero and $A_{T}$ becomes as

\begin{equation}
\left(A_{T}\right)_{F\rightarrow0}=\frac{\sum_{j}a_{j}\vert\alpha_{j}\beta_{j}^{\ast}\vert^{2}}{\sum_{j}\vert\alpha_{j}\beta_{j}^{\ast}\vert^{2}}=\langle A\rangle_{c}.
\end{equation}
On the contracy, using $gt^{2}\vert a_{i}-a_{j}\vert/2m\ll\sigma(t)$,
the overlap $\langle\phi_{i}(t)\vert\phi_{j}(t)\rangle$ approximately
equals one, and $A_{T}$ is reduced to
\begin{align}
\left(A_{T}\right)_{F\rightarrow1} & =\frac{\sum_{i}a_{i}\vert\alpha_{i}\beta_{i}\vert^{2}+\sum_{i\neq j}a_{i}\beta_{j}\beta_{i}^{\ast}\alpha_{i}\alpha_{j}^{\ast}}{\sum_{i}\vert\alpha_{i}\beta_{i}\vert^{2}+\sum_{i\neq j}\beta_{j}\beta_{i}^{\ast}\alpha_{i}\alpha_{j}^{\ast}}\nonumber \\
 & =\frac{\sum_{j}a_{j}\alpha_{j}\beta_{j}^{\ast}}{\sum_{j}\beta_{j}^{\ast}\alpha_{j}}=\langle A\rangle_{w}.\label{eq:33-1}
\end{align}
 From an experimental point of view, we obtain the above values of
the system observable $A$ by reading the position and momentum shifts
of the pointer. As given in the above subsection, the $\vert\Psi(t)\rangle$
{[}see Eq. (\ref{eq:18}){]} is the total system state of the our
system described by the Hamiltonian in Eq. (\ref{eq:9}) after the
time evolution. If we take a post-selection on it using the post-selected
state $\vert\psi_{f}\rangle$, then the unnormalized final state of
the pointer is given as
\begin{equation}
\vert\varXi(t)\rangle=\sum_{i}\beta_{i}^{\ast}\alpha_{i}\vert\phi_{i}(t)\rangle.
\end{equation}
 Using this final state provides the position and momentum shifts
of the pointer, and their expression are expressed as
\begin{align}
\delta x & =\frac{\langle\varXi(t)\vert x\vert\varXi(t)\rangle}{\langle\varXi(t)\vert\varXi(t)\rangle}-\langle\phi\vert x\vert\phi\rangle\nonumber \\
 & =\frac{1}{\sum_{i,j}\alpha_{i}\alpha_{j}^{\ast}\beta_{i}^{\ast}\beta_{j}F_{ij}}\{\sum_{i,j}\alpha_{i}\alpha_{j}^{\ast}\beta_{i}^{\ast}\beta_{j}{\textstyle \frac{gt^{2}(a_{i}+a_{j})}{4m}}F_{ij}\nonumber \\
 & +\!\!igt\sum_{i,j}\alpha_{i}\alpha_{j}^{\ast}\beta_{i}^{\ast}\beta_{j}({\textstyle \frac{t^{2}(a_{i}-a_{j})}{8\sigma^{2}m^{2}}}\!-\!\left(a_{i}-a_{j}\right)\sigma^{2}(t))F_{ij}\}\nonumber \\
 & =\frac{gt^{2}}{2m}Re(A_{T})+\frac{gt^{3}}{4\sigma^{2}m^{2}}Im(A_{T})-2gt\sigma^{2}(t)Im(A_{T}),\label{eq:35}
\end{align}
 and 
\begin{align}
\delta p & =\frac{\langle\varXi(t)\vert p\vert\varXi(t)\rangle}{\langle\varXi(t)\vert\varXi(t)\rangle}-\langle\phi\vert p\vert\phi\rangle\nonumber \\
 & =\frac{\sum_{i,j}\alpha_{i}\alpha_{j}^{\ast}\beta_{i}^{\ast}\beta_{j}\left[{\textstyle \frac{gt(a_{i}+a_{j})}{2}}-i{\textstyle \frac{gt^{2}(a_{i}-a_{j})}{8m\sigma^{2}}}\right]F_{ij}}{\sum_{i,j}\alpha_{i}\alpha_{j}^{\ast}\beta_{i}^{\ast}\beta_{j}F_{ij}}\nonumber \\
 & =gtRe(A_{T})-\frac{gt^{2}}{4\sigma^{2}m}Im(A_{T}),\label{eq:36-1}
\end{align}
respectively. If the transition factor $F$ approaches one, the shift
in the pointer's position and momentum behave as
\begin{align}
\left(\delta x\right)_{F\rightarrow1} & =\frac{gt^{2}}{2m}Re\left[\langle A\rangle_{w}\right]-\frac{gt^{3}+8gt\sigma^{4}m^{2}}{4\sigma^{2}m^{2}}Im\left[\langle A\rangle_{w}\right],\\
\left(\delta p\right)_{F\rightarrow1} & =gtRe\left[\langle A\rangle_{w}\right]-\frac{gt^{2}}{4\sigma^{2}m}Im\left[\langle A\rangle_{w}\right].
\end{align}
 On the other hand, if $F=0$ then Eqs.(\ref{eq:35}) and (\ref{eq:36-1})
reduce to
\begin{equation}
\left(\delta x\right)_{F\rightarrow0}=\frac{gt^{2}}{2m}\langle A\rangle_{c},
\end{equation}
and 
\begin{equation}
\left(\delta p\right)_{F\rightarrow0}=gt\langle A\rangle_{c}.
\end{equation}
Since the decohence factor $F$ is a continuous function, its two
extreme cases of can establish a relationship between weak and strong
measurements. As noticedin this work, both displacements in the weak
and strong measurement regimes do not coincide with the results obtained
by Josza \citep{PhysRevA.76.044103} and Turek \citep{TUREK2023128663},
owing to the Hamiltonian of our scheme. Since we aim to investigate
the mechanism behind the weak-to-strong measurement transition and
its relations with pointer-induced decoherence, we consider the pointer's
kinetic energy. However, in previous works, we only considered the
interaction of the Hamiltonian between the measured system and the
pointer. However, one interesting point of our scheme is that if particle
mass is assumed to be too heavy, the above displacements reproduce
the previous results, i.e., 
\begin{align}
\left(\delta x\right)_{F\rightarrow1,m\rightarrow\infty} & =-2gt\sigma^{2}Im\left[\langle A\rangle_{w}\right],\\
\left(\delta p\right)_{F\rightarrow1,m\rightarrow\infty} & =gtRe\left[\langle A\rangle_{w}\right],
\end{align}
 and 

\begin{align}
\left(\delta x\right)_{F\rightarrow0,m\rightarrow\infty} & =0,\\
\left(\delta p\right)_{F\rightarrow0,m\rightarrow\infty} & =gt\langle A\rangle_{c},
\end{align}
respectively. Most existing studies consider the interaction between
Hamiltonian $H_{I}$ to be in $g\text{\ensuremath{\hat{A}\otimes\hat{P}}}$
form, whereas in this work, $H_{I}=g\hat{A}\otimes\hat{X}$. Thus,
contrary to the previous results, in a weak measurement regime, the
position shift of the pointer is proportional to the imaginary part
of the weak value, and the momentum shift gives the real part of the
weak value. It is worth noting that the original paper of Aharonov
\citep{Aharonov1988HowTR} used the same interaction Hamiltonian as
in this paper. Fig.\ref{fig:2} highlights the above relations. The
next section provides a feasible example of the proposed scheme.

\section{\label{sec:3}A typical example--the Stern-Gerlach (SG) experiment }

The Stern-Gerlach (SG) experiment is a very important quantum measurement
model, which reflects the relationship between spin and spatial degrees
of freedom in atoms, and makes it possible to distinguishing different
spin states from spatial distributions. In the SG experiment, a silver
atom in the ground state with orbital angular momentum $L=0$ moves
along the $x$ direction and enters the non-uniform magnetic field
directed on the $z$-axis. This process is described by the Hamiltonian,
which can be written as 
\begin{align}
H & =\frac{p^{2}}{2m}-\mu B(x)\sigma_{z}.\label{eq:25}
\end{align}
Here, $m$ is the mass of the atom. If we take a linear approximation
$B(x)\approx\frac{\partial B}{\partial x}\vert_{x=0}x$, then the
above Hamiltonian becomes as
\begin{align}
H & =\left(\begin{array}{cc}
H_{+}\ \  & 0\\
0 & H_{-}
\end{array}\right),\label{eq:27}
\end{align}
where $H_{\pm}=\frac{p^{2}}{2m}\mp fx$ and $f=\mu\frac{\partial B}{\partial x}\vert_{x=0}x$.
If we assume that the initial system and pointer prepared to $\vert\psi_{i}\rangle=\cos\theta_{1}\vert\uparrow\rangle+e^{i\delta_{1}}\sin\theta_{1}\vert\downarrow\rangle$
and $\vert\phi\rangle$ {[}see Eq. (\ref{eq:11}){]}, the time evolution
of the total system is given by
\begin{equation}
\vert\Phi(t)\rangle=\cos\theta_{1}\vert\uparrow\rangle\vert\phi_{+}(t)\rangle+e^{i\delta}\sin\theta_{1}\vert\downarrow\rangle\vert\phi_{-}(t)\rangle.\label{eq:28}
\end{equation}
 Here, $\vert\phi_{\pm}(t)\rangle=e^{-iH_{\pm}t}\vert\phi\rangle$
and their explicit expression in position representation is obtained
as 

\begin{align}
\phi_{\pm}(x,t) & =\frac{\left(\sigma^{2}/2\pi\right)^{1/4}}{\sqrt{\sigma^{2}+\frac{it}{2m}}}e^{-i\theta^{\prime}(t)}e^{\mp iftx}\nonumber \\
 & \times\exp\left[-\frac{\left(x\pm\frac{ft^{2}}{2m}\right)^{2}\left(\sigma^{2}-\frac{it}{2m}\right)}{4\sigma^{4}+\frac{t^{2}}{m^{2}}}\right]\text{,}\label{eq:32}
\end{align}
with $\theta^{\prime}(t)=\frac{f^{2}t^{3}}{6m}$. The width of this
Gaussian wavepacket is $\sigma(t)$ with a central position at $x_{\pm}=\pm\frac{ft^{2}}{2m}$.
Every wavepacket has the same group velocity but propagates in opposite
directions, i.e., $\upsilon_{\pm}=\pm ft/m$. The expression presented
above infers that the central motion of wavepackets obeys the rule
of classical dynamics, i.e., an object with mass $m$ has acceleration
$f/m$ under the external force $f$. 

It can observed that the spatial degree of freedom and internal degree
(spin) of freedom of the silver atoms are entangled. Since the atoms
in different spin states $\left|{\rm \uparrow}\right\rangle $ and
$\left|\downarrow\right\rangle $ experience opposite forces, the
atom with the initial superposition of the two spin states will eventually
form two macroscopic distinguishable spots on the detection screen.
Once a particle is found at the position associated with the $\left|{\rm \uparrow}\right\rangle $
state, the state of the system's state is said to collapse to the
$\left|{\rm \uparrow}\right\rangle $ state, and vice versa.

The reduced density operator corresponding to the spin degrees of
freedom of the atom is given as
\begin{align}
\rho_{s}(t) & =Tr_{p}\left(\vert\Phi(t)\rangle\langle\Phi(t)\vert\right)\nonumber \\
 & =\cos^{2}\theta_{1}\vert\uparrow\rangle\langle\uparrow\vert+\sin^{2}\theta_{1}\vert\downarrow\rangle\langle\downarrow\vert\nonumber \\
 & +\frac{1}{2}e^{-i\delta_{1}}\sin2\theta_{1}\vert\uparrow\rangle\langle\downarrow\vert F^{\prime}(t)+h.c\label{eq:29}
\end{align}
 with $F^{\prime}(t)=\langle\phi_{+}(t)\vert\phi_{-}(t)\rangle$ and
its explicit expression reads as
\begin{align}
F^{\prime} & (t)=\exp\left[-\frac{\gamma^{2}}{8}-\frac{1}{32}\frac{t^{2}}{m^{2}\sigma^{4}}\gamma^{2}-2\sigma^{2}f^{2}t^{2}\right].\label{eq:30}
\end{align}
 where $\gamma^{\prime}=ft^{2}/m\sigma(t)$. This expression can be
obtained directly from Eq. (\ref{eq:21}) by substituting the $a_{i,j}=\text{\ensuremath{\pm1}}$
eigenvalues. Furthermore, if the post-selection with the system state
$\vert\psi_{f}\rangle=\cos\theta_{2}\vert\uparrow\rangle+e^{i\delta_{2}}\sin\theta_{2}\vert\downarrow\rangle$
is placed onto $\vert\Phi(t)\rangle$, the normalized final state
of the pointer in the position representation is obtained as 
\begin{equation}
\varOmega(x,t)=\frac{\cos\theta_{1}\cos\theta_{2}\phi_{+}(x,t)+\sin\theta_{1}\sin\theta_{2}e^{i(\delta_{1}-\delta_{2})}\phi_{-}(x,t)}{\beta},\label{eq:31}
\end{equation}
 Here, $\beta$ is the normalized coefficient, which is given by
\begin{align}
\beta^{2} & =cos^{2}\theta_{1}cos^{2}\theta_{2}+sin^{2}\theta_{1}sin^{2}\theta_{2}\nonumber \\
 & +\frac{1}{2}sin2\theta_{1}sin2\theta_{2}cos(\delta_{1}-\delta_{2})F^{\prime}(t).\label{eq:33}
\end{align}
 We obtain the explicit expressions of position and momentum shift
by using $\Omega(x,t)$, and the results expresssed as
\begin{align}
\delta x & =\frac{1}{\beta^{2}}\{\frac{ft^{2}}{2m}(cos^{2}\theta_{1}cos^{2}\theta_{2}-sin^{2}\theta_{1}sin^{2}\theta_{2})\nonumber \\
 & +\frac{ft^{3}+8ft\sigma^{4}m^{2}}{8\sigma^{2}m^{2}}sin2\theta_{1}sin2\theta_{2}\sin(\delta_{1}-\delta_{2})F^{\prime}(t)\},\label{eq:43}
\end{align}
 and 
\begin{align}
\delta p & =\frac{1}{\beta^{2}}\{ft(cos^{2}\theta_{1}cos^{2}\theta_{2}-sin^{2}\theta_{1}sin^{2}\theta_{2})\nonumber \\
 & +\frac{ft^{2}}{8\sigma^{2}m}sin2\theta_{1}sin2\theta_{2}\sin(\delta_{1}-\delta_{2})F^{\prime}(t)\},\label{eq:44}
\end{align}
 respectively.

Then, if $F^{\prime}\to0$, we consider the larger value of the coupling
strength parameter to know the position and momentum shifts of the
pointer in the post-selected strong measurement regime, which can
be written as
\begin{align}
\left(\delta x\right)_{F^{\prime}\rightarrow0} & =\!\!\frac{ft^{2}}{2m}\frac{cos^{2}\theta_{1}cos^{2}\theta_{2}-sin^{2}\theta_{1}sin^{2}\theta_{2}}{cos^{2}\theta_{1}cos^{2}\theta_{2}+sin^{2}\theta_{1}sin^{2}\theta_{2}}=\frac{ft^{2}}{2m}\left\langle \sigma_{z}\right\rangle _{c}.\label{eq:36}
\end{align}
 and 
\begin{equation}
\left(\delta p\right)_{F^{\prime}\rightarrow0}=ft\frac{cos^{2}\theta_{1}cos^{2}\theta_{2}-sin^{2}\theta_{1}sin^{2}\theta_{2}}{cos^{2}\theta_{1}cos^{2}\theta_{2}+sin^{2}\theta_{1}sin^{2}\theta_{2}}=ft\left\langle \sigma_{z}\right\rangle _{c}.\label{eq:46}
\end{equation}
Here, $\left\langle \sigma_{z}\right\rangle _{c}$ is the conditional
expectation value of observable $\sigma_{z}$, which is obtained in
a conditional strong measurement. 

Furthermore, if one wants to know the position shift formula for the
post-selected weak measurement regime, a limit $F^{\prime}\to1$ should
be taken for this extreme case. Then, the position and momentum shifts
become as
\begin{align}
\left(\delta x\right)_{F^{\prime}\rightarrow1} & =\!\!\frac{ft^{2}}{2m}Re\left[\left\langle \sigma_{z}\right\rangle _{w}\right]-\frac{ft^{3}+8ft\sigma^{4}m^{2}}{4\sigma^{2}m^{2}}Im\left[\left\langle \sigma_{z}\right\rangle _{w}\right],\label{eq:37}
\end{align}
and 
\begin{equation}
\left(\delta p\right)_{F^{\prime}\rightarrow1}=ftRe\left[\left\langle \sigma_{z}\right\rangle _{w}\right]-\frac{ft^{2}}{4\sigma^{2}m}Im\left[\left\langle \sigma_{z}\right\rangle _{w}\right].\label{eq:48}
\end{equation}
These are the general results of the SG experiment in the post-selected
weak measurement. However, in the dynamical evolution of the measurement
process, we usually assume the mass of the pointer to be too large
and do not consider the effects of the pointer caused by itself. In
this case, we omit the terms in the above results associated with
mass $m$, thereby recovering the typical displacements presented
in previous studies.

\section{\label{sec:4}Discussion and conclusion}

The experimental results reveal that whether we propose a general
model of ideal measurement or explain it through the specific SG experiment,
the decoherence factor obtained from the decoherence process and the
transition factor in the weak-to-strong measurement transition exhibit
the same mathematical form. This similarity raises the question of
whether the decoherence process of the measured system caused the
weak-to-strong measurement transition and weak measurement procedure
as well.

The decoherence factor given in Eq.(\ref{eq:21}) can be rewritten
as 
\begin{align}
F & =\exp\left[-\frac{1}{8}\frac{(\triangle x)^{2}}{\sigma^{2}(t)}-\frac{t^{2}}{32\sigma^{4}m^{2}}\frac{(\triangle x)^{2}}{\sigma^{2}(t)}-\frac{2m^{2}\sigma^{2}(\triangle x)^{2}}{t^{2}}\right].\label{eq:59}
\end{align}
 This factor also occurred in the weak-to-strong measurement transition
process. The value of $F$ depends on some parameters, including time
$t,$ mass of the atom $m$, coupling strength $g$, and atomic beam
width $\sigma$. Among these parameters, we can easily control the
time $t$ and coupling strength $g$. After the dynamical evolution,
the atomic beam width changed from $\sigma$ to $\sigma(t)=\sigma\left(1+\frac{t^{2}}{4m^{2}\sigma^{4}}\right)^{1/2}$.
Thus, during dynamic evolution, the wavepacket of the atomic beam
spreads in space. Decoherence arises from the interaction between
the measured system and the measuring apparatus during the quantum
measurement. Under the evolution of time, the system will change from
a superposition state that embodies quantum coherence to a mixed state.
Hence, utilizing a density matrix becomes essential to describe the
local system with greater relevance. However, providing the exact
decoherence time of our decoherence factor $F$ is impossible, but
we can discuss the two extreme cases with the time scale given in
$\sigma(t)$. If the dynamical evolution time $t$ is too long so
that $t\gg m\sigma^{2}$, then 
\begin{align}
F & \approx\exp\left[-\frac{g^{2}(a_{i}-a_{j})^{2}}{32\sigma^{2}m^{2}}t^{4}\right].\label{eq:60}
\end{align}
If $t$ or $g$ or both are large, this factor tends to zero, and
it can characterize the complete decoherence. In the context of decoherence,
the density matrix of the system's final state approaches this limit,
causing the off-diagonal elements that characterize coherence to become
zero. Taking the SG experiment as an example, when $F=0$, the system's
final state density matrix transforms into a completely mixed state
$\hat{\rho}_{s}(t)=cos^{2}\theta_{1}\left|\uparrow\right\rangle \left\langle \uparrow\right|+sin^{2}\theta_{1}\left|\downarrow\right\rangle \left\langle \downarrow\right|$.
Moreover, since the overlap is zero, indicating orthogonality between
$\vert\uparrow\rangle$ and $\langle\downarrow\vert$, and the information
about the apparatus's final state can be effectively distinguished.
Additionally, the decoherence discussed here arises from the strong
interaction between the system and the apparatus induced during the
measurement process. On the other hand, when \$F\$ approaches zero,
the measurement is considered strong in the weak-to-strong measurement
of the transition process. In this case, the displacement of the apparatus's
position is proportional to the conditional expectation value. When
the factor approaches zero, the prepared quantum system will collapse,
allowing us to differentiate the information about the apparatus's
final state effectively within a single measurement.

If we consider the very short time case, i.e., $t\ll m\sigma^{2}$,
then the decoherence factor Eq. (\ref{eq:59}) is reduced to
\begin{equation}
F\approx exp\left[-\frac{\sigma^{2}g^{2}\left(a_{i}-a_{j}\right)^{2}t^{2}}{2}\right]\approx1-\tau^{2}t^{2},
\end{equation}
 where $\tau=g\sigma(a_{i}-a_{j})/\sqrt{2}$. This kind of Gauss attenuation
can be considered a quantum Zero effect \citep{1990T}. In this process,
transitions between quantum states are inhibited by frequent state
measurements. The inhibition phenomena arises because the measurement
causes wave function collapse. If the time between measurements is
short enough, the wave function usually collapses back to the initial
state, and the main point of the post-selected weak measurement could
occur. In this case, since the factors on the off-diagonal elements
representing coherence tend to be one, the system's density matrix
is constant. As given in the SG experiment, the amount of overlap
quantifies the degree of interference based on the system. When $F=1$,
the reduced density matrix of the measured system after dynamical
evolution becomes $\rho_{s}=cos^{2}\theta_{1}\left|\uparrow\right\rangle \left\langle \uparrow\right|+sin^{2}\theta_{1}\left|\downarrow\right\rangle \left\langle \downarrow\right|+e^{-i\delta}cos\theta_{1}sin\theta_{1}\left|\uparrow\right\rangle \left\langle \downarrow\right|+e^{i\delta}cos\theta_{1}sin\theta_{1}\left|\downarrow\right\rangle \left\langle \uparrow\right|$,
as the initial state never touched.

Similarly, in the weak-to-strong measurement transition, when the
$F\to1$, the measurement is considered weak. In this process, the
position displacements and momentum of the pointer are proportional
to the imaginary and real parts of weak value. Since the interaction
strength between the system and the apparatus is weak in the weak-measurement
process, the shifts displayed on the dial of the pointer during a
single measurement are insufficient to achieve the desired measurement
outcome. However, a zero-effective process in enough short time or
weak coupling cases allows us to take multiple consecutive measurements
to obtain statistical results. Thus, we can confirm that controlling
the quantum decoherence is essential to performing the weak measurement.

This paper studied the relations between quantum decoherence and weak-to-strong
measurement transition. We observed that there exists a certain connection
between decoherence and weak-to-strong transitions, as they share
common features and exhibit analogous behaviors regarding these factors.
We presented the general expression of the decoherence factor in weak-to-strong
measurement transition by taking the pointer's Hamiltonian into account.
We found that in the pointer-based decoherence process, the mass $m$
of the pointer is highly important. Additionally, we noticed that,
in general, quantum measurement's dynamical evolution, whether weak
or strong, displacements of position and momentum of the pointer are
not zero. Specifically, in a weak measurement regime, both displacements
of position and momentum observables are proportional to the real
and imaginary parts of the weak value, respectively. In contrast,
in a strong measurement regime, they are both associated with conditional
expectation value. This result is not very common for the measurement
community. However, assuming the mass of this pointer is too heavy
to disturb the measurement result, the results presented in this paper
recovered the previous results. Overall, the proposed scheme can help
deepen the understanding of weak-to-strong measurement and clarify
the hiding mechanism behind the weak measurement theory.
\begin{acknowledgments}
This work was supported by the National Natural Science Foundation
of China (No. 12365005)
\end{acknowledgments}

\bibliographystyle{apsrev4-1}
\bibliography{reference}

\end{document}